\documentclass[twocolumn,showpacs,preprintnumbers,amsmath,amssymb]{revtex4}

\usepackage{graphicx}
\usepackage{dcolumn}
\usepackage{bm}

\begin{document}

\preprint{Physical Review B}

\title{Crystal structure and physical properties of $^{243}$AmPd$_{5}$Al$_{2}$}

\author{J.-C.~Griveau$^{1}$, K.~Gofryk$^{1,2}$, D. Bou\"{e}xi\`{e}re$^{1}$, E. Colineau$^{1}$, and J.~Rebizant$^{1}$}
\affiliation{$^{1}$European Commission, Joint Research Centre, Institute for Transuranium Elements, Postfach 2340,
Karlsruhe 76125, Germany\\ $^{2}$Los Alamos National Laboratory, Los Alamos,
New Mexico 87545, USA}

\date{\today}

\begin{abstract}
We report on the crystal structure, magnetic susceptibility, specific heat, electrical and thermoelectrical properties of
AmPd$_{5}$Al$_{2}$, the americium counterpart of the unconventional superconductor NpPd$_{5}$Al$_{2}$. AmPd$_{5}$Al$_{2}$
crystallizes in the  ZrNi$_{2}$Al$_{5}$-type of structure with lattice parameters: $a$~=~4.1298~\AA~and $c$~=~14.7925~\AA.
Magnetic measurements of AmPd$_{5}$Al$_{2}$ indicate a paramagnetic behavior with no hint of magnetic
ordering nor superconductivity down to 2~K. This aspect is directly related to its \emph{5f$^6$} electronic configuration with
J~=~0. The specific heat measurements confirm the non magnetic ground state of this compound. The low temperature electronic
specific heat $\gamma_{e}$~$\sim$~20~mJ~mol$^{-1}$K$^{-2}$ is clearly enhanced as compared to americium metal. All transport
measurements obtained point to a metallic behavior in AmPd$_{5}$Al$_{2}$.

\end{abstract}

\maketitle
\section{Introduction}
The discovery of unconventional superconductivity in transuranium based intermetallics such as PuCoGa$_{5}$\cite{1} and
PuRhGa$_{5}$\cite{2} has lead the scientific community to investigate the properties of numerous transuranium compounds,
especially at low temperatures. NpPd$_{5}$Al$_{2}$\cite{3} is the third transuranium and the first neptunium based
superconductor discovered within the last few years. Despite the fact that superconducting properties of
NpPd$_{5}$Al$_{2}$\cite{3,4,5} are analogous to \emph{4f} or \emph{5f} well studied Ce- or U-based heavy fermion
superconductors\cite{6} there are still many open questions that have not been resolved such as the pairing mechanism and the
symmetry of the order parameter. In the case of transuranium based superconductors several scenarios were proposed recently to
account for their superconductivity but the situation is still unclear [see Ref.~\onlinecite{1,bang,roberto}]. While
NpPd$_{5}$Al$_{2}$ is a heavy-fermion superconductor, the isostructural ThPd$_{5}$Al$_{2}$\cite{7} and
UPd$_{5}$Al$_{2}$\cite{8} exhibit a paramagnetic ground state and PuPd$_{5}$Al$_{2}$ orders antiferromagnetically below
5.6~K\cite{7}. The rare earth (RE) based compounds of the same structure such as CePd$_{5}$Al$_{2}$\cite{9} and
LuPd$_{5}$Al$_{2}$\cite{10} have also been examined and some of them present fascinating features such as pressure induced
unconventional superconductivity\cite{11}.

To explore further the \emph{f}-electron properties and the richness of behavior intrinsically related to this
ZrNi$_{2}$Al$_{5}$ structure, we have considered to substitute another transuranium element, namely americium. Americium
(Am$^{3+}$) presents a non magnetic ground state due to its \emph{5}f$^6$ electronic configuration (J~=~0). In the case of
americium metal, this implies a multitude of interesting properties such as superconductivity \cite{12} and an extreme
sensitivity to external parameters like pressure\cite{13} with a complex structural phase diagram \cite{14}. It suggests then
that an americium based compound in this structure should present interesting properties.

Here we report on the crystal structure and magnetic, transport and thermal properties of AmPd$_{5}$Al$_{2}$. It is worth to
note the challenge and the difficulties when working with americium: it is a highly radioactive element and only one isotope
allows low temperature studies ($^{243}$Am, t$_{1/2}$~=~7.38~x~10$^{3}$ years, $Q$=~6.3~$\mu$W~mg$^{-1}$). This isotope in metal form
at quantity required for sample preparation ($\sim$ 0.1~g scale) and physical property measurements is extremely rare. Such
difficulties have generally prevented studies of americium based intermetallics or compounds in the last decades especially
considering electrical transport properties (resistivity, magnetoresistance, thermopower) and specific heat at low temperature.
This study constitutes therefore a clear advance on trans-plutonium basic properties at low temperatures. This should be added
to the few ones reported\cite{15,16} and would help to understand the nature of the unconventional superconductivity in
NpPd$_{5}$Al$_{2}$, especially the magnetic fluctuation scenario related to 5\emph{f}-electrons.

\section{Experimental}
The polycrystalline samples were prepared by arc melting stoichiometric amounts of the pure metals components. Starting
materials were used in the form of 4N~Pd pellet, 5N~Al wire as supplied by A. D. Mackay Inc. and 99.85~\%~Am metal($^{243}$Am
isotope\cite{Mueller}) produced originally by thermo reduction process \cite{Spirlet} with 3N purity. Other actinides (Np, Pu)
are present at ppm level\cite{RhoAm} in Am after long term storage ($\sim$10 years). They do not contribute significantly to
basic properties but decrease purity level with time. Due to the contamination risk generated by the radiotoxicity of
americium, all operations of preparation and encapsulation have been performed in gloveboxes under inert N$_{2}$ atmosphere.
Moreover the self radiation and the self damage induced by $\alpha$ disintegration constrain us to work very rapidly (within
days). This aspect is even more drastic at low temperature (T~$<$10~K). The arc melting was performed under a high purity argon
atmosphere on a water-cooled copper hearth, using a Zr alloy as an oxygen/nitrogen getter. The arc melted buttons were turned
over and melted three times in order to ensure homogeneity. The weight losses after melting were smaller than 0.2~\%. The
magnetic properties were determined using a Quantum Design (QD) MPMS-7 device in the temperature range 2-300~K and in magnetic
fields up to 7~T. The heat capacity, electrical resistivity and Hall effect were measured from 2.5 to 300~K with a QD PPMS-14
device up to 14~T on polycrystals. The thermoelectric power was measured from 3 to 300~K in a home-made setup using pure copper
as a reference material. For transport properties, all measurements have been determined by a four~probe~DC technique voltage
measurement on a polycrystal sample polished on 2 opposite faces to present a flat surface with typical size
1x0.4x0.2~mm$^{3}$. Electrical resistivity measurement has been performed with the applied current $I$~=~5~mA along the flat
surface and the voltage $V$ was extracted parallel to the current $I$ while the applied magnetic field  $B$ was perpendicular to
the flat surface for magnetoresistivity determination. When applying negative field at fixed temperature, no parasitic Hall
effect was observed for magnetoresistivity. The Hall resistance (R$_{H}$) was determined by voltage measurements V$_{H}$ under
field alternatively at $+14$ and $-14~T$. Field response V$_{H}$(B) at fixed temperatures has been measured to confirm results
obtained when ramping in temperature. Excepted crystallographic structure determination, no orientation information related
could be extracted from the basic properties observed.

\section{Crystallographic structure}

\begin{figure}[t!]
\includegraphics[width=0.4\textwidth]{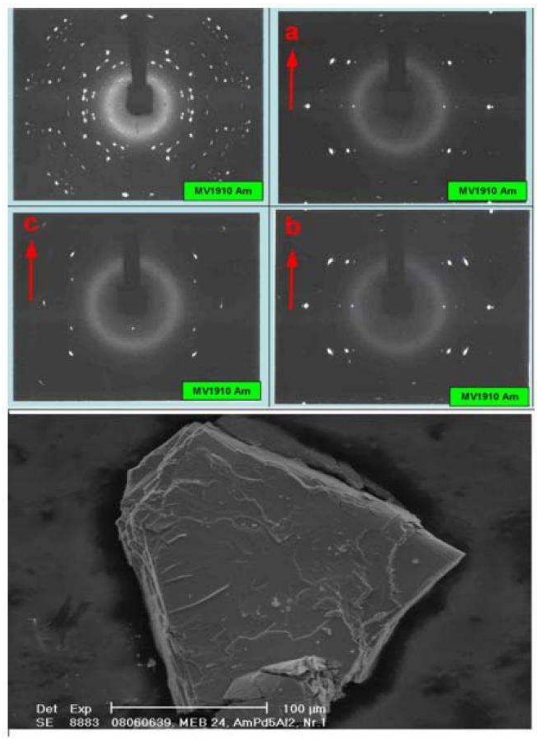}
\begin{centering}
\caption{(Color online) X-ray single crystal diffraction pattern obtained for a single crystal for a random orientation and
along the main directions a, b and c. Bottom, energy dispersive x-ray (EDX) analysis performed on a polycrystal showing the
homogeneity of the material.}
\end{centering}
\end{figure}

\begin{figure}[t!]
\includegraphics[width=0.5\textwidth]{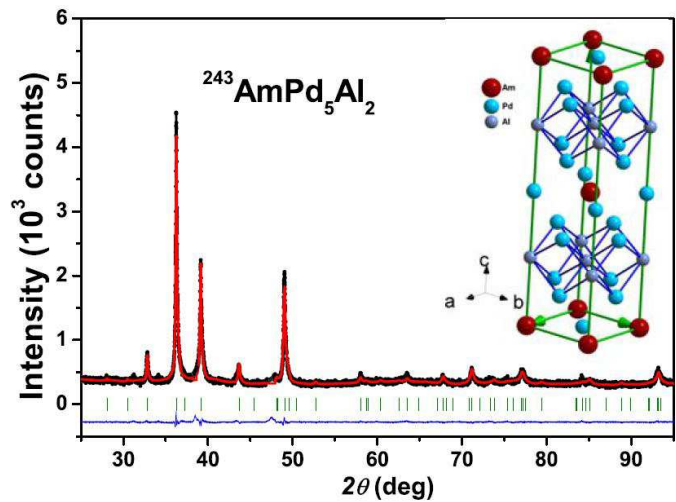}
\begin{centering}
\caption{(Color online) X-ray powder diffraction pattern ($\lambda$~=~1.54059~\AA) recorded for $^{243}$AmPd$_{5}$Al$_{2}$. The
solid red line through experimental points is the Rietveld refinement profile calculated for tetragonal AmPd$_{5}$Al$_{2}$. The
blue line corresponds to the difference between measured and calculated Rietveld refinement. The crystal structure with the
ZrNi$_{2}$Al$_{5}$ unit cell and the crystallographic parameters $a$~=~4.1298(9)~\AA~and $c$~=~14.792(4)~\AA~are presented.}
\end{centering}
\end{figure}

\begin{table}[b!]
\caption{Crystallographic data for $^{243}$AmPd$_{5}$Al$_{2}$ obtained from a single crystal.} \label{table:1}
\newcommand{\m}{\hphantom{$-$}}
\newcommand{\cc}[1]{\multicolumn{1}{c}{#1}}
\renewcommand{\tabcolsep}{1pc} 
\renewcommand{\arraystretch}{1.2} 
\begin{tabular}{@{}ll}
\\
\hline
Composition                       &       $^{243}$AmPd$_{5}$Al$_{2}$\\
Space~Group                      &       I4/mmm~(No.~139)\\
Lattice~Parameters~(~\AA)        &       a~=~4.1298(9)\\
                                 &       c~=~14.792(4)\\
Cell~Volume (~\AA$^{3}$)         &       252.29\\
Formula~units~per~cell           &       Z~=~2 \\
Formula~Mass                     &       829.07\\
Calculated~density~(g/cm$^{3}$)  &       10.91\\
Crystal~size~(mm$^{3}$)               &       ~0.03~x~0.20~x~0.05\\
Radiation                        &       MoK$_{\alpha}$ ($\lambda$~=~0.71073~\AA)\\
Scans up to 2$\theta$            &       70~$^o$\\
Linear absorption coefficient    &        32.00~mm$^{-1}$\\
Total number of reflections     &       1008\\
                                &       Unique:~201~(R$_{F}$ = 0.118)\\
Reflections  with F$_{o}>4\sigma$(F$_{o}$)  &     154\\
Goodness~of~fit                 &       1.134\\
Conventional~residual~R        &          0.0772    (F~$>~4\sigma$)\\
\hline
\end{tabular}\\[2pt]
\end{table}

\begin{table}[t!] \caption{Crystallographic parameters for $^{243}$AmPd$_{5}$Al$_{2}$ obtained by single crystal x-ray diffraction} \label{table:2}
\newcommand{\m}{\hphantom{$-$}}
\newcommand{\cc}[1]{\multicolumn{2}{c}{#1}}
\renewcommand{\tabcolsep}{1pc} 
\renewcommand{\arraystretch}{1.2} 
\begin{tabular}{@{}llllll}
\hline
Atom           & site & x & y & z & U$_{eq}$\\
\hline
Am             & 2a & 0   & 0   & 0         & 0.0060\\
Pd1            & 8g & 0.5 & 0   & 0.1453(2) & 0.0069\\
Pd2            & 2b & 0.5 & 0.5 & 0         & 0.0056\\
Al             & 4e & 0   & 0   & 0.2545(20)& 0.0106\\
\hline
\end{tabular}\\[2pt]
The structure was refined with anisotropic displacement parameters for all atoms. The last column contains the
equivalent isotropic U values (~\AA$^{2}$).
\end{table}

Small single crystals of typical size 50x50x20~$\mu$m$^{3}$ suitable for crystal structure determination were mechanically
extracted from the button and mounted inside a capillary for single crystal x-ray diffraction (Fig.~1 top). They were examined
on an Enraf-Nonius CAD-4 diffractometer with the graphite monochromatized MoK$\alpha$-radiation. The crystal structure was
solved by the direct method using SHELX97\cite{20} and the data processing using the WinGX package\cite{21} are reported in
Table~I. The atomic coordinates obtained are presented in Table~II. The crystal structure was refined from the single-crystal
x-ray data and was corrected for Lorentz and polarization effects. The results confirm the tetragonal unit cell. Samples were
then examined by powder x-ray diffraction using a Bruker D8 diffractometer with the monochromated CuK$\alpha_{1}$ radiation
($\lambda$=1.54059~\AA) equipped with a Vantec detector. The powder diffraction pattern (Fig.~2) was analyzed by a Rietveld
profile refinement method\cite{18} using the WinplotR-Fullprof program\cite{19}. Results are similar to those determined by
x-ray single crystal diffraction confirming the structure, the crystallographic parameters and good homogeneity of the
polycrystals. Finally, the phase composition was determined by energy dispersive x-ray (EDX) analysis performed on a Philips
XL40 scanning electron microscope (SEM). The microprobe analysis indicates good homogeneity (Fig.~1 bottom) for the samples and
a single phase with stoichiometry close to 1:5:1.70 (Am$_{0.13}$Pd$_{0.65}$Al$_{0.22}$). This slight deficiency in aluminium on
the bulk could be explained by the difficulty to integrate the Al signal vs. Am. In conclusion, AmPd$_{5}$Al$_{2}$ adopts a
tetragonal ZrNi$_{2}$Al$_{5}$-type of structure (s.g. \textit{I}4/\textit{mmm}) like NpPd$_{5}$Al$_{2}$\cite{3} with lattice
parameters $a$~=~4.1298(9)~\AA~and $c$~=~14.792(4)~\AA.

\section{Magnetic properties}

\begin{figure}[b!]
\includegraphics[width=0.5\textwidth]{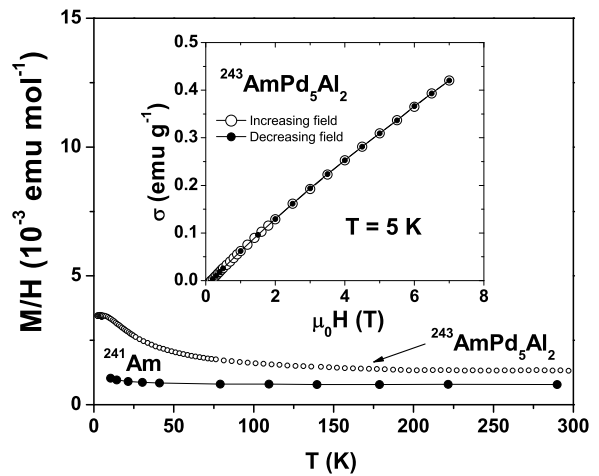}
\caption{Temperature dependence of the magnetic susceptibility M/H of $^{243}$AmPd$_{5}$Al$_{2}$ for H~=~70~kOe. The magnetic
susceptibility reported by \textit{Kanellakopoulos et al.}\cite{22} for Am metal is also plotted on the graph. Inset: Magnetization
$\sigma$ measured at 5~K for increasing and decreasing magnetic field H. The value is very small even at the highest applied magnetic field ($\sim$~0.4~emu~g$^{-1}$).}
\end{figure}

The temperature dependence of the magnetic susceptibility of polycrystalline AmPd$_{5}$Al$_{2}$ is presented in Fig.~3. The
compound shows a paramagnetic behavior with almost no temperature dependence down to 50~K. Below, an upturn leads to a slightly
enhanced magnetic susceptibility at low temperature with $\chi_{2K}$~$\sim$~3.5x10$^{-3}$~emu~mol~$^{-1}$. A similar but weaker
behavior has been reported for the magnetic susceptibility of americium metal\cite{22} and is presented on the same figure for
comparison. The field dependence of the magnetization ($\sigma$) taken at 5~K (see inset in Fig.~3) indicates a paramagnetic
response. The non magnetic nature of AmPd$_{5}$Al$_{2}$ is expected from the ground state of americium (5f$^{6}$). In LS
coupling the six 5\textit{f}electrons of Am$^{3+}$ will occupy the J~=~0 state. In this case, due to the proximity of the first
excited state J~=~1 the susceptibility will mainly be governed by a Van Vleck term. On the other hand in $jj$ model the six
electrons will occupy $j$=5/2 sub-shell separated by spin-orbit coupling from $j$=7/2 sub-level. So, the susceptibility coming
from a Pauli exchange reinforced term and an orbital contribution is rather anticipated. However, the exchange interactions are
still present leading to intermediate coupling of the 5\textit{f}-states. As has been shown in Am an intermediate coupling is
very close to the $jj$ limit so only a small 5\textit{f}-electron occupation is observed in $j$=7/2
sub-band\cite{kotliar,moore}. Therefore, the enhanced susceptibility values obtained for AmPd$_{5}$Al$_{2}$ compared to
americium metal may come from a stronger Van Vleck contribution and/or Pauli paramagnetism. The presence of Pu and more
specifically Np atoms created during the long term storage could be considered but can not explain quantitatively the behavior
at low temperature. They can be present in the material, participating to a relative local disorder in the lattice without
signatures on the XRD powder diffraction patterns or EDX measurements. No sign of superconductivity nor magnetic order has been
observed down to 2~K in AmPd$_{5}$Al$_{2}$.

\section{Transport properties}
\subsection{Electrical resistivity}

The electrical resistivity and magnetoresistance of polycrystalline samples of AmPd$_{5}$Al$_{2}$ are presented in Fig.~4. The
overall behavior of the electrical resistivity of AmPd$_{5}$Al$_{2}$ is clearly metallic with a relatively high value of the
residual resistivity $\rho_{0}$ and a low value of the residual resistivity ratio (RRR~=~1.35). The electrical resistivity
value is small and point to the absence of magnetic scattering phenomena such as Kondo or RKKY interactions. It is interesting
to note the similarity of behavior for all \emph{$4f$} and\emph{ $5f$} counterparts (RE,An)Pd$_{5}$Al$_{2}$ with electrical
resistivity $\rho_{300K}$ around 15-30~$\mu\Omega~cm$ \cite{5,7,9}. Only NpPd$_{5}$Al$_{2}$ presents a higher electrical
resistivity at room temperature $\sim$~90~$\mu\Omega~cm$ \cite{3,4}.

Assuming the validity of Matthiesen's rule the resistivity of a non-magnetic metallic compound should follow the
Bloch-Gr\"{u}neisen-Mott relation \cite{24,25}:
\begin{equation}
\rho(T)~=~\rho_{0}~+~4R\Theta_R(\frac{T}{\Theta_R})^{5}\int_0^{\frac{\Theta_R}{T}}\frac{x^{5}dx}{(e^{x}-1)(1-e^{-x})}~-~KT^{3}
\label{DebyeRho}
\end{equation}
where $\Theta_R$ is the Debye temperature obtained by resistivity and $R$ is a constant, whereas the third term $KT^{3}$
describes interband electron scattering processes on the s-d bands in the case of transition metal alloys and s-f bands in the
case of $f$ metals. The magnitude of the constant K and its sign depend on the density of states at the Fermi level. The
adjustment of the electrical resistivity curve by a LSQ fit as shown in Fig.~4 according to Eq. (1) gives the following
results: residual resistivity $\rho_{0}$~=~17.0~$\mu$$\Omega~cm$, phonon scattering term
$R$~=~$2.372$x10$^{-2}$~$\mu$$\Omega~cm~K^{-1}$, Debye-temperature $\Theta_R$~=~209~K and scattering coefficient of the
conduction electrons into a narrow d band near the Fermi level $K$~=~$2.978$x10$^{-8}$~$\mu$$\Omega$~cm~K$^{-3}$. This value of
Debye temperature is roughly of the same order of magnitude as obtained by heat capacity (200-300 K) but slightly reduced. A
second fit using the Debye temperature obtained by heat capacity measurements ($\Theta_D$~=~295~K, see IV)  is presented on the
same figure Fig.~4. The values are then $\rho_{0}$~=~17.1~$\mu$$\Omega~cm$, $R$~=~$2.50$x10$^{-2}$~$\mu$$\Omega~cm~K^{-1}$, and
$K$~=~$5.18$x10$^{-8}$~$\mu$$\Omega$~cm~K$^{-3}$. This fit does not reproduce well the overall shape of electrical resistivity
especially at low temperature. The discrepancy between the two fits can be explained by 2 origins.

On one side, the different phonons branches contributes differently to the 2 properties (heat capacity and electronic
transport). The electrical conductivity ($\sigma$~$\sim$~1/$\rho$) is predominantly limited by the interaction between
conduction electrons and longitudinal phonons, whereas the low temperature specific heat is dominated by transverse
phonons\cite{Enss}. This is clearly reinforced in the case of anisotropic structures presenting highly selective modes.

On the other side, the RRR can be reduced by the disorder created by the $^{243}$Am decay. We suggest to consider an increased
interaction at low temperature of some defects induced by self disintegration of $^{243}$Am atoms creating Frenkel pairs in the
lattice. These defects appear quite fast despite the short delay between synthesis and measurement (within a week) and play a
much stronger role than the presence of Np and Pu atoms created during long term storage that should not contribute
significantly to the transport properties as for the magnetization. They could lead to an "impurity effect" with time, reducing
the RRR as their effect is even more visible at low temperature when phonon modes are reduced. This phenomenon has been clearly
observed in intermetallic systems especially in the case of superconductors\cite{Jutier}.

We get the same value for the residual resistivity $\rho_{0}$ for both adjustments but slightly enhanced values for $R$ and
$K$ contants for the $\Theta_R$~=~209~K case. The $R$ constant is of the same order of magnitude than non magnetic
$5f$\cite{Troc} or localized/mixed valence $4f$ intermetallics $\sim~10^{-2}$~$\mu$$\Omega~cm~K^{-1}$ such as YbTIn$_{5}$
(T=Rh, Ir)\cite{Bukowski} and YbAl$_{2}$\cite{Nowatari} but $K$ constant is clearly reduced by one or two orders of magnitude
($10^{-8}$$\mu$$\Omega$~cm~K$^{-3}$ for AmPd$_{5}$Al$_{2}$) in comparison to UTGa$_{5}$(T=Co, Rh) for instance
($\sim~10^{-6}~$$\mu$$\Omega$~cm~K$^{-3}$ for URhGa$_{5}$)\cite{Wrawyk}. As the value of $K$ as well as its sign depend on the
density of states at the Fermi level, this illustrates the localized aspect of the material and the absence of
5\textit{f}-electrons at the Fermi level.

\begin{figure}[t!]
\includegraphics[width=0.5\textwidth]{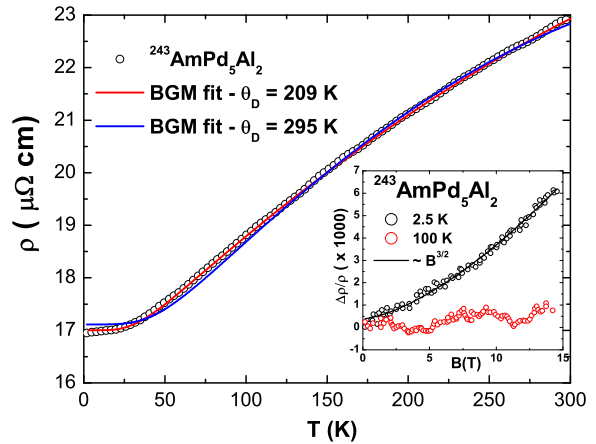}
\caption{(Color online)The temperature dependence of the electrical
resistivity of AmPd$_{5}$Al$_{2}$.  Two adjustments by the
Bloch-Gruneisen Model are presented (with $\Theta_D$~ as free
parameter and =~295~K). Inset: magnetoresistivity taken at 2.5 and
100~K up to 14~T. A B$^{3/2}$ regime is visible at 2.5~K.}
\end{figure}

The inset in Fig.~4 presents the magnetoresistivity ($MR$) data $\Delta \rho$/$\rho$~$\times$~1000 taken at 2.5 and
100~K, respectively. As may be seen from the figure we observe, in AmPd$_{5}$Al$_{2}$, an extremely small positive
magnetoresistivity ($<$~1~$\%$). At 2.5~K the field dependence of the resistivity may be well described by a power law
behavior B$^{3/2}$ (see the solid line in the inset of Fig.~3a). It is slightly different from $\rho(B)$ observed in pure
simple metals where resistivity grows quadratically with field \cite{25}. This can be related to the large anisotropy
of the cristallographic structure as reported for other counterparts of the 1:5:2 family\cite{3,10} and by the
"impurity effect" \cite{Graf}. This impurity effect is noticeable on magnetotransport because all other contributions are
negligible: the \emph{5f}-electrons are localized and are not participating to the bonding. The small positive values of
$MR$ points to closed \emph{5f} shell characteristics. Only remaining \emph{s,p,d} electrons can interact and lead to a
scattering effect noticeable on the metallic-like shape of the material and on the weak magnetotransport properties. To
estimate the electronic carriers qualitatively and quantitatively we have performed Hall and Seebeck effect measurements.

\subsection{Hall and Seebeck effects}

The temperature dependence of the Hall coefficient $R_{H}$ measured in a magnetic field of 14~T is shown in Fig.~5. $R_{H}$ is
positive up to 70~K where it changes sign to negative. Unlike simple metals, $R_{H}$ of AmPd$_{5}$Al$_{2}$ is strongly
temperature dependent. This could suggest, in AmPd$_{5}$Al$_{2}$ the presence of a complex electronic structure with multiple
electron and hole bands with different temperature variations of carrier concentrations and mobilities. The higher value of
Hall effect at low temperatures could come from the reduction of the carrier mobility when going down to 0~K. Above 100~K the
$R_{H}$ is of the order of -2$\times$10$^{-10}~$m$^{3}$~C$^{-1}$, which corresponds in a one band model to the effective electron
concentration $n_{Hall}$~$\sim$~$3.2$$\times$10$^{22}$~cm$^{-3}$. It agrees well with the thermoelectric data (see below).

\begin{figure}[t!]
\includegraphics[width=0.5\textwidth]{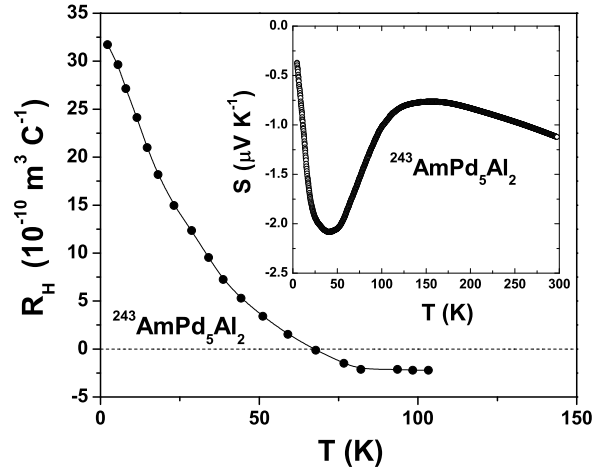}
\caption{The temperature dependence of the Hall effect and thermoelectric power (inset) of AmPd$_{5}$Al$_{2}$.}
\end{figure}

The inset of Fig.~5 shows the temperature dependence of the thermoelectric power of AmPd$_{5}$Al$_{2}$. It is worth noting
that, up to our best knowledge, it is the first measurement of the Seebeck coefficient of a Am-based system. The thermopower is
negative in the whole temperature range suggesting the dominant role of electrons in the electrical and heat conduction. The
low-temperature dependence of the thermoelectric power $S(T)$ shows a clear extremum at $T_{ext}$ = 45~K. Then, with decreasing
temperature $S$ decreases down to zero when T$\rightarrow$0~K, as expected. This extremum of the thermopower can be related to
the phonon drag effect, where the position is a measure of the Debye temperature [$\Theta_{D} \approx$ $T_{ext}$/5 (see
Ref.\onlinecite{Thermopower law})], which we estimate for AmPd$_{5}$Al$_{2}$ to be of about 230~K. It is of the same order as
$\Theta_{D}$ values estimated from the electrical resistivity and specific heat measurements (see below). Above 150 K the
thermopower is roughly proportional to the temperature, hence indicating that the main contribution to $S(T)$ comes from
diffusion of the carriers due to the applied temperature gradient. This mechanism is generally expressed as\cite{Thermopower
law}:

\begin{equation}
S(T) = \frac{k^2_{\rm B}\pi^2T}{3eE_{\rm F}}.
\end{equation}

Within a single-band model the value of the thermoelectric power measured at room temperature (-1.1~$\mu$V~K$^{-1}$) implies the Fermi
energy and effective carrier concentrations to be about $E_{F}$~=~6.5~eV and
$n_{Seebeck}$~$\sim$~$7.5$$\times$10$^{22}$~cm$^{-3}$, respectively. The value of $n$ is very close to the one obtained from the
Hall effect measurements and is also similar to $n$ observed in usual metals such as Al or Cu\cite{25,26}. However, taking into
account the simplicity of the model applied, the estimated carrier concentration could account for the upper limit of carrier
concentration in AmPd$_{5}$Al$_{2}$. Altogether, it strongly suggests the absence of $\emph{5f}$ electrons in the electronic
properties: only $\emph{s-p-d}$ carriers participate to the bonding while \emph{5f}-electrons are well localized in
AmPd$_{5}$Al$_{2}$.

\section{Heat capacity}

The specific heat measurements of AmPd$_{5}$Al$_{2}$ are presented
on Fig.~6. The absence of magnetic order and/or superconductivity in
this system down to 3~K is clearly confirmed. Near room temperature
$C_{p}$ has a value of about 180~J~mol$^{-1}$K~$^{-1}$ which is
slightly lower than the Dulong-Petit limit, i.e., $C_{p}$=3$rR$=199
J/mol~K, where $r$ is the number of atoms per molecule and $R$ is
the gas constant. The low temperature part of the specific heat
allows to determine the electronic contribution $\gamma_e$ and Debye
temperature $\theta_D$ by a linear fit of
C$_{p}$/T$\sim$~$\gamma_e$+$\beta$T$^{2}$, with $\beta$ related to
$\theta_D$ (see inset of Fig.~6). For AmPd$_{5}$Al$_{2}$ we get
$\gamma_e$=19~mJ~mol$^{-1}~$.K$^{-2}$ and $\theta_D$=~295~K. The value
of $\theta_D$ is very close to the other non magnetic counterparts
of the family, $\theta_D$~=~290~K for ThPd$_{5}$Al$_{2}$ and 310~K
for LuPd$_{5}$Al$_{2}$. The value of $\gamma_e$ obtained is
relatively enhanced especially when compared to americium
metal\cite{27} or to other classical metals. However this value is
much smaller than all other parent compound of the
ZrNi$_{2}$Al$_{5}$ structure (325~mJ~mol$^{-1}$~K$^{-2}$ for
NpPd$_{5}$Al$_{2}$\cite{5} and 61~mJ~mol$^{-1}$~K$^{-2}$ for
PuPd$_{5}$Al$_{2}$\cite{7}) and is rather close to the one for
ThPd$_{5}$Al$_{2}$ (4.4~mJ~mol$^{-1}$~K$^{-2}$)as this latter one
does not present any $\emph{5f}$ contribution. It suggests the
presence of well localized 5\textit{f}-electrons in
AmPd$_{5}$Al$_{2}$. The slight enhancement of $\gamma_e$ could come
from the $\emph{5d-6d}$ exchange correlations between Pd and Am
atoms. However, further measurements using synchrotron techniques as
magnetic x-ray circular dichroism are needed to conclude on this
point.

\begin{figure}[t!]
\includegraphics[width=0.5\textwidth]{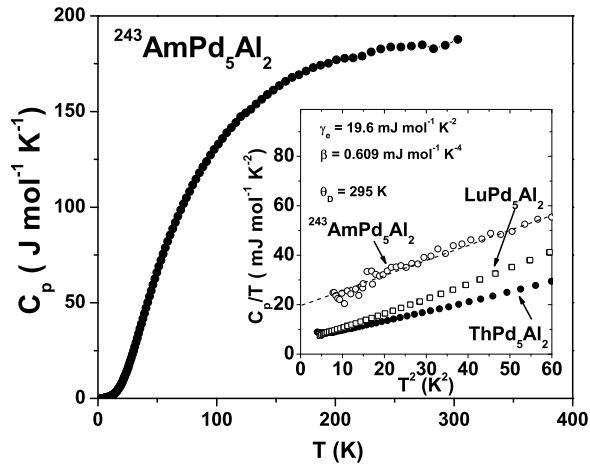}
\caption{The temperature dependence of the heat capacity of AmPd$_{5}$Al$_{2}$. Inset: low temperature specific heat of
AmPd$_{5}$Al$_{2}$ plotted as C$_{p}$/T vs T$^{2}$ (see text).}
\end{figure}

\section{Discussion and conclusion}

In conclusion, the Am-based intermetallic compound
AmPd$_{5}$Al$_{2}$ crystallizes, like other members of the
AnPd$_{5}$Al$_{2}$ family in the tetragonal ZrNi$_{2}$Al$_{5}$-type
of structure (s.g. \textit{I}4/\textit{mmm}) with lattice parameters
$a$~=~4.1298(9)~\AA~and $c$~=~14.793(4)~\AA, as determined by single
crystal studies. The magnetic measurements revealed that
AmPd$_{5}$Al$_{2}$ shows a temperature independent paramagnetic
behavior, enhanced compared to Am metal. In agreement with the
electronic configuration of americium, it does not show any hint of
a magnetic nor superconducting signature down to 2~K. The non
magnetic ground state, governed by the J=0 configuration, is also
supported by specific heat measurements. The electrical resistivity,
Hall effect and thermoelectric power are characteristic of a good
metallic system with the carrier concentration of the order of
10$^{22}$cm$^{-3}$. This clearly point to the presence in
AmPd$_{5}$Al$_{2}$ of well localized 5\textit{f}-electrons.
Moreover, the absence of superconductivity in this system strongly
emphasizes the importance of magnetic interactions as a possible
medium of the unconventional superconductivity in
NpPd$_{5}$Al$_{2}$. Electronic structure calculations would be of
great interest to determine the position of $\emph{5f}$ states
versus Fermi energy in AmPd$_{5}$Al$_{2}$ and compare it to the
other members of the (An,RE)Pd$_{5}$Al$_{2}$ family.

\begin{acknowledgments}
We are grateful to H. Thiele for technical assistance for the EDX analysis. K.G. acknowledges the European Commission
for support in the frame of the "training and Mobility of Researchers" program.
\end{acknowledgments}

\end{document}